\documentclass[aps,prl,twocolumn,showpacs, superscriptaddress]{revtex4}
\usepackage{graphicx,amssymb,amsfonts,amsmath}

\begin{document}

\title{
Nonequilibrium quantum criticality in open electronic systems
}
\author{Aditi Mitra}
\affiliation{Department of Physics, University of Toronto, 60 St.
George Street, Toronto, Ontario, M5S 1A7 Canada}
\author{So Takei}
\affiliation{Department of Physics, University of Toronto, 60 St.
George Street, Toronto, ON, Canada}
\author{Yong Baek Kim}
\affiliation{Department of Physics, University of Toronto, 60 St.
George Street, Toronto, ON, Canada}
\author{A. J. Millis}
\affiliation{Department of Physics, Columbia University, 538 W.
120th Street, New York, NY 10027 USA}
\date{\today}


\begin{abstract}
A theory is presented  of  quantum criticality in open (coupled to reservoirs)
itinerant electron magnets,
with nonequilibrium drive provided by current flow across the system.
Both departures from equilibrium at conventional (equilibrium) quantum
critical points and the physics of phase transitions induced by the
nonequilibrium drive are treated. Nonequilibrium-induced phase transitions
are found to have the same leading critical behavior as conventional thermal phase transitions.
\end{abstract}


\maketitle

A central issue in condensed matter physics
is the behavior of systems as one tunes  parameters (for example pressure,
or magnetic field) so as to change the
symmetries characterizing the
ground state \cite{Hertz76,Millis93,Sondhi97,Sachdev99}.
The parameter values at which the
ground
state symmetries  change (for example, from ferromagnetic metal to paramagnetic metal) define a
quantum phase transition point (quantum critical point).
At quantum phase transitions, spatial and temporal fluctuations are coupled, so that
continuous quantum phase  transitions in equilibrium systems are typically
described by critical theories involving an effective dimensionality
$d_{\rm eff}$ greater than the spatial dimensionality $d$.

While equilibrium quantum phase transitions have been extensively
studied, the generalization to nonequilibrium conditions raises a
largely open class of questions. Nonlinear transport near a
superconductor-insulator phase transition \cite{philip,sondhigreen}
and a ferromagnetic transition driven by current flow in a closed
one-dimensional system \cite{feldman} have been studied; however, a
general systematic understanding is lacking.

\begin{figure}
\includegraphics[totalheight=6cm,width=6cm]{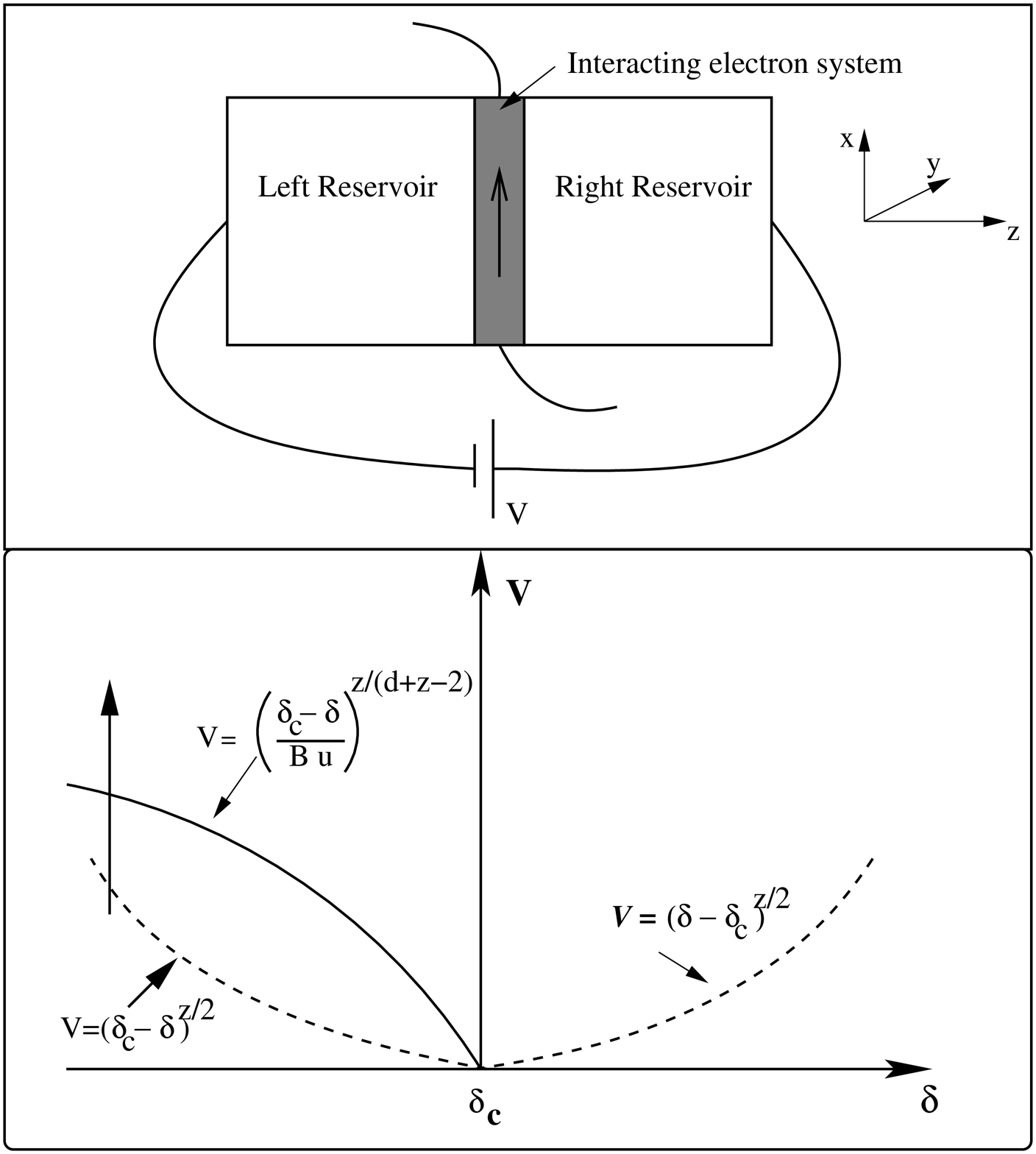}
\caption{Upper panel: Schematic view of systems studied:
interacting electron system coupled to two leads.
Lower panel: Schematic phase diagram in a plane of
equilibrium distance from  criticality, $\delta$, and departure from equilibrium $V$.
The quantum critical point $\delta_c$ separates the long range ordered
($\delta < \delta_c$) and disordered ($\delta > \delta_c$) phases.
%
The solid line denotes a nonequilibrium phase transition.
The dashed curves indicate the crossover
from low $V$ essentially equilibrium physics to the
higher-$V$ nonequilibrium-dominated regime.
}
\label{schem}
\end{figure}

In this paper we  formulate a  theory of nonequilibrium
quantum criticality in itinerant electron systems coupled to reservoirs (c.f. upper panel of
Fig.~\ref{schem}) with which particles
may be exchanged. Nonequilibrium is imposed by differences between reservoirs; our systems
are therefore subject to a time-independent drive,
and are not characterized by any conserved quantities.
A generic phase diagram
is shown in the lower panel of Fig \ref{schem}: we take a system
which at temperature $T=0$ may be tuned through an equilibrium
quantum critical point by varying a parameter $\delta$ through a
critical value $\delta_c$ and determine the changes induced by a
nonequilibrium drive (generically denoted as $V$). Of particular
interest is the transition generated by $V$ if the $V=0$ system is
ordered (vertical arrow in lower panel of Fig.~\ref{schem}).

Analysis of nonequilibrium systems proceeds from the time dependent density matrix
${\hat \rho}(t)$ defined in terms of a Hamiltonian ${\hat H}$ and
an initial condition ${\hat \rho}(t_{init})$ via
${\hat \rho}(t)=e^{-i{\hat H}(t-t_{init})}{\hat \rho}(t_{init})e^{i{\hat H}(t-t_{init})}$.
The open systems we consider possess a well defined long-time  state defined by
a density matrix ${\hat \rho}_{SS}$  independent of
the initial condition ${\hat \rho}(t_{init})$.
%
We use the usual Suzuki-Trotter breakup to express the problem as a
two-time contour functional integral
\cite{Keldysh63,Kamenev04,Mitra05a}, Hubbard-Stratonovich techniques
to decouple the interaction terms by introducing auxiliary fields
which we interpret as order parameter fluctuations, and integrate
over the electronic degrees of freedom. The result is that the
physics of the steady state system is expressed in terms of a
generating functional, ${\cal Z}$, of source fields,
$\eta$,\cite{Kamenev04,Mitra05a}
\begin{eqnarray}
&{\cal Z}(\eta)=\int {\cal D}[m_i,m_f] \rho_{SS,\Lambda}\left(m_i,m_f\right)\int^{'} {\cal D}
[m_+(t),m_-(t)]\nonumber \\
&e^{S_K[\{m_+,m_-,\eta\}]}
\label{zdef}
\end{eqnarray}
Here $m_{\pm}$ are the fluctuating order parameter fields of
interest and $\Lambda$ is a short-distance cutoff. $\int^{'} {\cal
D}[\{m_+(t),m_-(t)\}]$ denotes an integral over all paths in
function space beginning at $m_i$ on the $+$ contour at $t=0$ and
ending at $m_f$ at $t=0$ on the $-$ contour. The contributions of
paths
with given endpoints are weighted by the steady-state reduced density matrix
$\rho_\Lambda[\{m_i(x),m_f(x)\}]$, whose diagonal elements describe the probability that at
an instant of time the long wavelength components of the order parameter field take the configuration
$m(x)$. Eq \ref{zdef} is the generalization to nonequilibrium systems of the
partition function used to treat criticality in equilibrium systems. The important differences
are the two time contours and the presence of  $\rho_{SS}$
which obeys a kinetic equation  determined
\cite{Mitra05a} from the requirement
that correlation functions calculated from Eq \ref{zdef} are causal and finite.
Eq \ref{zdef} may be analysed by the renormalization group method
of integrating out modes near the cutoff and
rescaling.

We apply the formalism to the case of a two dimensional itinerant magnet
placed between two noninteracting leads,
with current flowing across the system
(cf Fig \ref{schem}). We model the system  by the Hamiltonian
\begin{eqnarray}
H&=&H_{layer}+H_{mix}+H_{leads} \\
H_{layer}&=&\sum_{i,\delta,\sigma}t_{\delta}c^\dagger_{i+\delta,\sigma}c_{i,\sigma}+H_{int} \\
H_{mix}&=&\sum_{i,k,\sigma,b=L,R}\left(V^k_{b}c^\dagger_{i,\sigma}a_{i,k,\sigma,b}+h.c.\right)
\end{eqnarray}
where the combination of the band structure implied by $t$ and the interactions
$H_{int}$ is such that the isolated layer is ferromagnetic.
The lead electrons are described by operators $a$
and have free fermion correlators $\left<a^\dagger_{b}a_b\right>=f_b$ with $f_b$ a fermi function
with lead-dependent chemical potential $\mu_b$.
For $V^k_b\neq 0$ neither the number of electrons nor the magnetization
in the layer are conserved.

We assume, as is usual in studies of quantum critical phenomena
\cite{Hertz76,Millis93},
that the electronic propagators
and susceptibilities in the layer take the usual fermi liquid form
and treat the interactions by a perturbative renormalization group. The presence of the leads
implies that the Greens functions describing the propagation of electrons in the interacting layer
are
\begin{eqnarray}
&G^{R}(p,\omega)=\frac{1}{\omega-\varepsilon_p-\Sigma^R(p,\omega)}=\left(G^A\right)^* \\
&G^K(p,\omega)=\frac{\Sigma^K(p,\omega)}{\left(\omega-\varepsilon_p - Re\Sigma^R(p,\omega)\right)^2+
\left|Im \Sigma^R(p,\omega)\right|^2}
\end{eqnarray}
where $p$ is a two dimensional momentum within the layer,
 $Im \Sigma^R=\sum_a\Gamma_a$ with $\Gamma_a(p,\omega)$ being the
rate at which electrons escape from the active layer into the lead $a$ and
$\Sigma^K=-2 i\sum_{a=L,R}\Gamma_a(p,\omega)\left(1-2f(\omega-\mu_a)\right)$
determines the distribution function imposed by coupling to reservoirs.
We shall focus on excitation energies less than $\Gamma_a$ and momenta less than
$Im\Sigma^R/v_F$ where nonconservation due to escape into the leads is dominant.
The analysis sketched above then leads to a generating function of the form of
Eq \ref{zdef}, with
$S_K=S^{(2)}+S^{(4)}+...$
where
\begin{eqnarray}
&S^{(2)} =
-i\int dt \int dt^{\prime} \int d^d r \int d^d r^{\prime}
\begin{pmatrix}{m_{cl}(t,r)} & m_{q}(t,r)
\end{pmatrix} \nonumber \\
&\begin{pmatrix} 0& \left[ \chi^{-1} \right]^{A}
\\
\left[ \chi^{-1} \right]^{R} &  \Pi^{K}
\end{pmatrix}
\begin{pmatrix}m_{cl}(t^{\prime},r^{\prime})\\
m_{q}(t^{\prime},r^{\prime})\end{pmatrix}
\label{Zk1}
\end{eqnarray}
Here $m_{q} = \frac{m_--m_+}{2},m_{cl}= \frac{m_- + m_+}{2}$,
the ellipsis denotes terms of higher than fourth order in $m$ and the fourth order
term $S^{(4)}$ will be presented and discussed below.
The quadratic-level inverse propagators are:
\begin{eqnarray}
&\left[\chi^{-1}\right]^R(q,\Omega)=\left(\left[\chi^{-1}\right]^A\right)^*=
\delta+ \frac{-i\Omega}{\gamma}+\xi_0^2q^2 +\ldots
\label{chiR}\\
&\Pi^{K}(q,\Omega)=-2i  \sum_{a b}
\coth\frac{\Omega +\mu_{a}-\mu_{b}}{2T}\frac{(\Omega +\mu_{\alpha}-\mu_{\beta})}
{\gamma^{ab}}
\label{PiK}
\end{eqnarray}
The key quantity is $\Pi^{K}$.
In  equilibrium systems at $T=0$, $\Pi^K(t)$ vanishes  at long times
(as a power law for itinerant-electron models); however at $T\neq 0$ and
(for all of the models we have studied) out of equilibrium
$\Pi^{K}(t\rightarrow\infty)\neq 0$.
Mathematically, $\Pi_K$ acts as a mass for the quantum fluctuations.
If $T$ or $V =|\mu_a - \mu_b| \neq 0$, quantum fluctuations are
gapped and at long times the theory is classical.  The
$\left(\chi^{-1}\right)^{(R,A)}$ describe non-conserved (because of
the leads) overdamped magnetization fluctuations.
$(\gamma_{ab})^{-1}=\left<\Gamma_a(p,\Omega=0)
\Gamma_b(p,\Omega=0)/\Gamma^3(p,\Omega=0)\right> |_{FS}$ are fermi
surface averaged decay rates, and $\gamma^{-1} = \sum_{a b = L,R}
\gamma_{ab}^{-1}$. The ``mass'' (distance from criticality) $\delta$
depends on the interaction, layer density of states and coupling to
the leads. The overdamped dynamics implies that even at $V=T=0$ the
momentum conjugate to $m(q,t)$ is logarithmically large, so that the
dc fluctuations are  essentially classical. The density matrix is
then easily obtained from the generating function using the
techniques of \cite{Kamenev04}. We find, up to corrections of
${\cal{O}}(V^2/\Gamma^2)$ in the argument of the exponential,
\begin{equation}
\rho[m_i(k),m_f(k)] \sim \delta_{m_i,m_f} \exp\left[-\frac{2
Re\left[\chi^{-1}(k)\right]^{R}|m_i(k)|^2} {i\gamma \Pi^K(\Omega=0)}
\right] \label{densitymatrix}
\end{equation}

After calculation, we find that the leading nonlinearity is
\begin{eqnarray}
S^{(4)} = -i\int (d\{k\}) \sum_{i=1...4}u_i m_q^im_{cl}^{4-i}
\end{eqnarray}
Here the $u_i$ are interaction functions which depend on the momenta
and frequency $\{k\}$ of all of the fields. The level broadening due
to the leads means any space-dependence may be neglected. In an
isolated system with  Hamiltonian dynamics one would have $u_1=u_3$
independent of frequency and $u_{2,4}=0$. The coupling to a
reservoir means that the interactions are retarded and that
$u_{2,4}\neq 0$. The limit of $u_{1,3}$ as all momenta and
frequencies tend to zero is real and positive, so to obtain the
leading long-wavelength, low energy behavior we may treat $u_{1,3}$
as constants.  However, at $T=V=0$, $u_{2,4}\rightarrow 0$ as the
external frequencies are set to zero so the frequency dependence
must be considered. This is somewhat involved, but the case
important for subsequent considerations is when the two quantum
fields carry a frequency $\pm \Omega$; in this case at $V=0$,
$u_2\rightarrow u_2' \Omega \coth\frac{\Omega}{2T}$, while at $T=0,
V\neq0$, $u_2 \rightarrow u^{\prime \prime}_2 |V|$, with real and
positive $iu_2',iu^{\prime \prime}_2$.

We now  formulate a renormalization group treatment, following along
the usual lines \cite{Sachdev99, Hertz76, Millis93}: we choose a
$b>1$ and in Eq \ref{Zk1} integrate out those fluctuations with
momenta between $\Lambda$ and $\Lambda/b$, treating the interactions
perturbatively. One technical remark is needed: causality implies
$\left<m_{q}m_{q}\right>$ correlator vanish identically at all
times. In a theory with a frequency cutoff, care must be exercised
to ensure that the cutoff does not violate the causality
requirements. We find it simpler to work with a theory with a
momentum cutoff but no frequency cutoff, so that we eliminate all
frequencies for each removed momentum mode. We next rescale
$q,\omega,T,V$ in order to keep the cutoff, coefficients of $q^2$
and $i \Omega$, and arguments of $\Pi^K$ invariant; thus $q
\rightarrow \frac{q^{\prime}}{b}$, $\Omega \rightarrow
\frac{\Omega^{\prime}}{b^z}$, $T,V \rightarrow
\frac{T^{\prime},V^{\prime}}{b^z}$. Observe that under this
rescaling the density matrix preserves the form Eq
\ref{densitymatrix}. The result is a change in the mass $\delta$
\begin{eqnarray}
&\frac{d \delta}{d\ln b}=
2  \delta + 3 u_1(b) g \label{rgb}
\end{eqnarray}
A similar renormalization coming from $u_2$ preserves the form of
the $\Pi^K$ term but changes the coefficient. We interpret
this as a finite renormalization of the broadening parameter $\gamma$
and do not consider it further.
The mode elimination leads also
to a renormalization of the interactions (note: $\bar{u}_{2,4} = i u_{2,4}$,
$\epsilon = 4 - d - z$)
\begin{eqnarray}
&\frac{du_1}{d\ln b} = \epsilon u_1 - 18 u_1^2 (f^{KR} + f^{KA})
+ 12 u_1 \bar{u}_2 f^{RA}& \label{rg1}\\
&\frac{d\bar{u}_2}{d\ln b} = \epsilon \bar{u}_2  - 2 \bar{u}_2 \left[ 15 u_1 (f^{KR} + f^{KA})
- 2 \bar{u}_2 f^{RA} \right]&
\nonumber \\
&+ 18 u_1 \left[u_1 f^{KK}-2 u_3f^{RA}\right]&
\label{rg2}\\
&\frac{d(u_1-u_3)}{d\ln b} = \epsilon (u_1 - u_3) - 12 u_1 \left[6 u_4 f^{RA} - u_2 f^{KK}
\right] + \ldots \, \,\, & \label{rg3} \\
&\frac{d\bar{u}_4}{d\ln b} = \epsilon \bar{u}_4  - 6 \bar{u}_2 u_3 (f^{KR} + f^{KA}) + \ldots
&   \label{rg4}
\end{eqnarray}

Here $g=\frac{K_d}{2}\int \frac{d\omega}{2\pi} \tilde{\chi}_K$;
$f^{ab} = \frac{K_d}{4}\int \frac{d\omega}{2\pi}\tilde{\chi}^a
\tilde{\chi}^b$
where $\tilde{\chi}^a= \tilde{\chi}^a(\Lambda,\omega) $ is one of
$\chi^R,\chi^A$, $-i \chi^K= i\Pi^K \chi^R \chi^A$, and $K_d = \int \frac{d^d q}{(2\pi)^d}\delta
(q-\Lambda)$ (in our discussions we will set $\Lambda=1$).
Note that the initial values are $u_1 - u_3 = {\cal O} (T^2, V^2)$,
$\bar{u}_2, \bar{u}_4 = {\cal O} (T,V)$, $f^{KK} = 2f^{RA} + {\cal O }(T^2, V^2)$.
The functions
$f,g$ depend on $\delta(T,V)$.
When $\delta,T,V \rightarrow 0$, $f^{ab}$ and $g$ tend to constant values
of order unity. For $T,V > 1$
but $\delta \ll 1$,
$f^{KK}\sim V^2, T^2$, $(f^{KR}+ f^{KA}), g \sim V,T$ and
$f^{RA} \sim 1$.
When $\delta \sim 1$ scaling stops.

We
solve the scaling equations starting from the
physically relevant initial conditions $\delta, T, V \ll 1$, and taking $d=z=2$
as appropriate for a thin magnetic layer. To leading nontrivial order in
$T,V$ we need to retain only the first terms after the $\epsilon$ term in
Eqns~\ref{rg1},~\ref{rg2},~\ref{rg3},~\ref{rg4}. We integrate these equations
to obtain $\delta(b) = e^{2 \ln b} \left[\delta_0 + 3 g \int_0^{\ln b}
dx u_1(e^x) e^{-2 x} \right] $ where $u_{1} (b) = \frac{u_{10}}{1 + f_1 u_{10}  \ln b}$
, $f_1 = 18 (f^{KR} + f^{KA})$ and the subscript $0$ denotes initial conditions.
We integrate up to scales where
$\delta \sim 1$ (initial conditions corresponding to region below
dashed line in Fig \ref{schem}) or,  if $\delta$ remains small,
$max(V,T) \sim 1$. For the latter case, we denote the value of
$\delta$ at the crossover scale by $r$.
We focus on the interesting
regime $-1 \ll r \leq 0$ corresponding to criticality or to a $T=0$ ordered state
very near to the critical point (scaling trajectory depicted by
vertical arrow in Fig \ref{schem}). We henceforth set $T=0$.

At the crossover scale the interactions are all small by powers of
logarithms. In the crossover region $V\sim 1$ the expressions are
complicated. In the classical region $V  \gg 1$, the functions $f,g$
acquire the $V$ dependence noted above. Rewriting the scaling
equations in terms of $g = \bar{g}V, f^{KK} = \bar{f}^{KK}V^2,
f^{KR}+ f^{KA}= V(\bar{f}^{KR}+ {\bar f}^{KA}),
f^{RA}=\bar{f}^{RA}$,$v_1 = u_1 V$, $v_2 = \bar{u}_2$, $v_3 =
\frac{u_3}{V}$ and $v_4 = \frac{\bar{u}_4}{V^2}$ leads to
\begin{eqnarray}
&\frac{d\delta}{d\ln b} = 2 \delta + 3 v_1 \bar{g}  \\
&\frac{dv_1}{d\ln b} = 2 v_1 - 18 v_1^2 (\bar{f}^{KR}+ \bar{f}^{KA} )
+ 12 v_1 v_2 \bar{f}^{RA} \label{rg1a} \\
&\frac{d v_2}{d\ln b} = 18 v_1^2 \bar{f}^{KK} - 30 v_1 v_2 (\bar{f}^{KR}+ \bar{f}^{KA} )
\nonumber \\
&+ 4 v_2^2 \bar{f}^{RA}  - 36 v_1 v_3 \bar{f}^{RA}  \\
&\frac{d v_3}{d\ln b} = -2 v_3 + \ldots;\,\, \frac{d v_4}{d\ln b} = - 4 v_4 + \ldots
\end{eqnarray}
Thus in the classical regime $v_3$ and $v_4$ vanish rapidly, $v_1$
grows and $v_2$ reaches a fixed point. The effective theory becomes
quadratic in $m_q$ which may be integrated out \cite{Kamenev04}. The
result is a theory for the fluctuations of the classical component
of the magnetization field in the presence of a Gaussian,
delta-correlated noise determined by the nonequilibrium drive:
\begin{eqnarray}
-\frac{1}{\gamma}
\frac{\partial m_{cl}}{\partial t} &=&
(\delta - \xi_0^2 \nabla^2 + v_{1} m_{cl}^2) m_{cl}
+ \xi \nonumber \\
\langle \xi(x,t)\xi(x^{\prime},t^{\prime})\rangle
&=&\delta(x - x^{\prime}) \delta(t - t^{\prime})\frac{2T_{\rm eff}}{\gamma}
\label{noise}
\end{eqnarray}
with $T_{\rm eff}=V\frac{\gamma}{\gamma_{LR}}$ (we
have used a standard transformation \cite{Risken} to eliminate a coupling,
generated by $v_2$, between the noise and the classical field). All parameters
acquire finite renormalizations (not explicitly denoted) from the scaling process.
Eq \ref{noise} is identical to that used in the standard analysis of Model A
\cite{Hohenberg77} relaxational dynamics, except that $T_{\rm eff}$
appears instead of temperature in the noise correlator.
We therefore conclude that the voltage-driven
transition is in the same universality class as the usual thermal 2d Ising
transition, and more generally that in this model, as far as universal quantities
are concerned voltage acts  as a temperature.  From Eq \ref{noise}
we may extract
experimental consequences:
near the critical point, magnetic correlation length $\xi$ is
$\xi^{-2}= \delta \sim \frac{V}{\ln1/V}$ and the in-plane
$2d$ resistivity arising from the scattering of electron with
critical fluctuations
is $\rho(V) \sim V^{3/2}$.

We next extend the analysis to the $O(3)$
symmetric (Heisenberg) case. As in the equilibrium situation \cite{Hertz76,Millis93}
the physics of the disordered and quantum-classical crossover regimes
is only weakly dependent on spin symmetry.
Differences appear in the ``renormalized classical'' regime
corresponding to adding a weak nonequilibrium drive to an ordered
state. In this regime the procedure leading to Eq.~\ref{noise} gives
a nonlinear stochastic equation describing fluctuations of the
magnetization amplitude and precession of its direction. Here we
focus on the most important special case, namely precession of small
amplitude, low frequency, long-wavelength fluctuations of the
magnetization direction about a state assumed to possess long ranged
order directed along ${\hat z}$ and we denote the spin-gap by
$\Delta$. The important degrees of freedom are those transverse to
the ordering direction. We find that at scales $t>1/\gamma $ and $L>
(v_F/\gamma)$ these are described by
\begin{eqnarray}
&\left(\frac{a_{xx}}{\gamma }+
{\hat z} \times \frac{a_{xy}\Delta}{\gamma^2}\right)\frac{\partial{\vec m}}{\partial t}
-\left(b_{xx}-\frac{b_{xy}\Delta V }{\gamma \gamma_{LR}}{\hat z} \times\right) \xi_0^2\nabla^2 {\vec m}&
={\vec \xi} \nonumber \\
&&
\label{LLG}
\end{eqnarray}

${\vec \xi}$ in Eq.~\ref{LLG} is a fluctuating noise field whose components are independent
and correlated according to Eq \ref{noise} but with $\gamma$ replaced by $\gamma/a_{xx}$.
The $a,b$ coefficients are numbers of order unity.
The subscripts  indicate whether they arise
from the $xx$ or $xy$ terms in the retarded/advanced susceptibilities.
The term involving $a_{xy}$ gives the Landau-Lifshitz-Gilbert  spin precession
and is explicitly proportional to the spin gap $\Delta$ so is small near
the quantum critical point.
The term involving $a_{xx}$ expresses  the damping due
to coupling to the leads, remains nonvanishing as $\Delta \rightarrow 0$
and is  the dominant time derivative term.
Solving Eq \ref{LLG} in the rotating frame leads to
\begin{eqnarray}
&\langle m_+(q,t) m_-(q^{\prime} t^{\prime}) \rangle =
 \frac{T_{\rm{eff}} a_{xx}
\delta(q + q^{\prime})e^{-D_{\rm{eff}}q^2 \xi_0^2|t-t^{\prime}| -i \omega_{\rm{eff}}(t-t^{\prime})}
}{(a_{xx}b_{xx} - a_{xy} b_{xy}\frac{\Delta^2 V}{\gamma^2 \gamma_{LR}})q^2\xi_0^2}
\label{langsol}
\end{eqnarray}

Eq.~\ref{langsol} shows that the mean square magnetization
fluctuations diverge as $\frac{1}{q^2}$. This divergence signals the
instability of the ordered state by the voltage-induced decoherence
in precise analogy to the usual $2d$ thermal case. The oscillatory
term in Eq \ref{langsol} expresses the usual spin precession with
precession frequency $\omega_{\rm{eff}}=(\frac{a_{xy}b_{xx}\gamma^2
+ a_{xx}b_{xy}V\frac{\gamma^2}{ \gamma_{LR}}} {a_{xx}^2 \gamma^2 +
a_{xy}^2 \Delta^2} )\Delta q^2 \xi_0^2 $ shifted from the
equilibrium result
by an amount proportional to $V$  due to spin accumulation effects
at the interface of the magnetized layer and leads \cite{ExpTorque}.
The decaying term in Eq.~\ref{langsol} expresses the damping due to
coupling to leads $D_{\rm{eff}}=\frac{(a_{xx} b_{xx} \gamma^3 -
a_{xy} b_{xy}\Delta^2 V\frac{\gamma}{\gamma_{LR}})} {a_{xx}^2
\gamma^2 + a_{xy}^2 \Delta^2} $. If $ a_{xx} b_{xx} < a_{xy} b_{xy}
\frac{\Delta^2 V }{\gamma_{LR}\gamma^2}$, Eq \ref{LLG} supports
modes which grow exponentially with time leading to the spin-torque
instability recently discussed \cite{brouwer}. However, where the
present theory applies ($\frac{\Delta, V}{\gamma}\ll 1$) there is no
instability. Calculation reveals that obtaining a nonzero $b_{xy}$
also requires an energy-dependent asymmetry between the leads
($\Gamma_L(\varepsilon_1)\Gamma_R(\varepsilon_2)-
\Gamma_L(\varepsilon_2) \Gamma_R(\varepsilon_1) \neq 0$).

We have presented a theory for nonequilibrium phase transitions in
an itinerant-electron system coupled to external reservoirs. We
provide a precise mapping onto an effective classical theory which
demonstrates that the leading effect of the nonequilibrium drive is
to generate an effective temperature and hence a transition in the
standard Wilson-Fisher thermal universality class.
Nonequilibrium-induced breaking of time reversal and inversion
symmetries and the creation of a coherently precessing
(``spin-torque") state appear only at the level of subleading
corrections. The techniques introduced here can be applied to
important open problems such as systems where the drive couples
linearly to the order parameter \cite{philip,sondhigreen}, driven
Bose condensates \cite{Greiner02}, and the closed system, conserved
order parameter work of \cite{feldman}.

{\it Acknowledgements:}
This work was supported by NSF-DMR-0431350 (AJM) and NSERC,
CRC, CIAR, KRF-2005-070-C00044 (AM,ST,YBK)

{\it Note Added}: Shortly after our manuscript was submitted  a study of the
closely related problem of nonlinear transport at a quantum critical
point appeared \cite{Green2}.

\end{document}